\documentclass[draftcls,11pt,onecolumn]{IEEEtran}
\usepackage{eurosym}
\usepackage{cite,algorithm,array,url}
\usepackage{graphicx,colortbl}
\usepackage{subcaption}
\usepackage{amsfonts,times,amsmath,amssymb,amsthm}
\usepackage{multirow,multicol,booktabs}
\usepackage{amsmath}
\usepackage{algorithm}
\usepackage{algpseudocode}

\usepackage{pgf}

\usepackage[acronym, toc]{glossaries}
\newacronym{uas}{UAS}{Unmanned Aerial System}
\newacronym{eo}{EO}{Electro-Optical}
\newacronym{uav}{UAV}{Unmanned Aerial Vehicle}
\newacronym{rf}{RF}{Radio Frequency}
\newacronym{isr}{ISR}{Intelligence, Surveillance and Reconnaissance}
\newacronym[\glslongpluralkey={Targets Of Interest}]{toi}{TOI}{Target Of Interest}
\newacronym{toiw}{TOIW}{Target Of Interest Warning}
\newacronym{kcf}{KCF}{Kalman Consensus Filter}
\newacronym{kf}{KF}{Kalman Filter}
\newacronym{ber}{BER}{Bit Error Rate}
\newacronym{dht}{DHT}{Distributed Hypothesis Testing}
\newacronym{jpda}{JPDA}{Joint Probabilistic Data Association}
\newacronym{mtic}{MTIC}{Multi Target Information Consensus}
\newacronym{ntp}{NTP}{Network Time Protocol}
\newacronym{cbaa}{CBAA}{Consensus-Based Auction Algorithm}
\newacronym{cbba}{CBBA}{Consensus-Based Bundle Algorithm}
\newacronym{srnp}{SRNP}{Self Routing Network Protocol}
\newacronym{pca}{PCA}{Principal Components Analysis}
\newacronym{svm}{SVM}{Support Vector Machine}
\newacronym{ftms}{FTMS}{Freeway Traffic Monitoring Systems}
\newacronym{gmm}{GMM}{Gaussian Mixture Model}
\newacronym{hmm}{HMM}{Hidden Markov Model}
\newacronym{fis}{FIS}{Fuzzy Inference System}
\newacronym{fov}{FOV}{Field Of View}
\newacronym{smc}{SMC}{Statistical Model Checking}
\newacronym{gt}{GT}{Ground Truth}
\newacronym{hsi}{HSI}{HyperSpectral Imaging}
\newacronym{ml-nn}{ML-NN}{Multi-Layered Neural Network}
\newacronym{nn}{NN}{Neural Network}
\newacronym{cnn}{CNN}{Convolutional Neural Network}
\newacronym{rnn}{RNN}{Recurrent Neural Network}
\newacronym{ids}{IDS}{Intrusion Detection System}
\newacronym{os}{OS}{Operating System}
\newacronym{t2ta}{T2TA}{Track-to-Track Association}
\newacronym{mta}{MTA}{Measurement-to-Track Association}
\newacronym{kcif}{KCIF}{Kalman Consensus Information Filter}
\newacronym{nid-kcif}{NID-KCIF}{Naive Information Driven Kalman Consesus Information Filter}
\newacronym{did-kcif}{DID-KCIF}{Decoupled Information Driven Kalman Consesus Information Filter}
\newacronym{mse}{MSE}{Mean Squared Error}
\newacronym{rmse}{RMSE}{Root Mean Squared Error}
\newacronym{kb}{KB}{Knowledge-Based}
\newacronym{lb}{LB}{Learning-Based}
\newacronym{ab}{AB}{Average-Based}
\newacronym{gmti}{GMTI}{Ground Moving Target Indication}
\newacronym{fhwa}{FHWA}{Federal Highway Administration}
\newacronym{ngsim}{NGSIM}{Next Generation SIMulation}
\newacronym{sa}{SA}{Situational Awareness}
\newacronym[\glslongpluralkey={Gaussian Processes}]{gp}{GP}{Gaussian Process}
\newacronym{mf}{MF}{Membership Function}
\newacronym{sd}{SD}{Standard Deviation}
\newacronym{lti}{LTI}{Linear Time-Invariant}
\newacronym{lrt}{LRT}{Likelihood-Ratio Test}
\newacronym{elrt}{ELRT}{Extended Likelihood-Ratio Test}
\newacronym{elr}{ELR}{Extended Likelihood-Ratio}
\newacronym{mle}{MLE}{Maximum Likelihood Estimators}
\newacronym{rhs}{RHS}{Right-Hand Side}
\newacronym{lhs}{LHS}{Left-Hand Side}
\newacronym{cdf}{CDF}{Cumulative Distribution Function}
\newacronym[\glsshortpluralkey={regexes}]{regex}{regex}{Regular Expression}
\newacronym{dfa}{DFA}{Deterministic Finite Automata}
\newacronym{dm}{DM}{Driving Mode}
\newacronym[\glsshortpluralkey={UAVs/UASs},\glslongpluralkey={Unmanned Aerial Vehicles/Systems}]{uavuas}{UAV/UAS}{Unmanned Aerial Vehicle/System}
\newacronym{cdtt}{CDTT}{Consensus-Based Distributed Target Tracking}
\newacronym{icf}{ICF}{Information-Weighted Consensus Filter}
\newacronym{map}{MAP}{Maximum A Posteriori}
\newacronym{bn}{BN}{Bayesian Network}
\newacronym{sia}{SIA}{Stochastic, Indecomposable and Aperiodic}
\newacronym{tsk}{TSK}{Takagi-Sugeno-Kang}
\newacronym{sds}{SDS}{Sleepiness Detection System}
\newacronym{mrf}{MRF}{Markov Random Field}
\newacronym{hdp-hmm}{HDP-HMM}{Hierarchical Dirichlet Process-Hidden Markov Model}
\newacronym{hdp-hsmm}{HDP-HSMM}{Hierarchical Dirichlet Process-Hidden Semi Markov Model}
\newacronym{ugv}{UGV}{Unmanned Ground Vehicle}
\newacronym{us}{US}{Unmanned System}
\newacronym{mas}{MAS}{Multi-Agent System}
\newacronym{mrs}{MRS}{Multi-Robot Systems}

\newcommand{\eqn}[1]{\textrm{Eqn.}~(\ref{#1})}
\newcommand{\eqns}[1]{\textrm{Eqns.}~(\ref{#1})}
\newcommand{\eqnref}[1]{(\ref{#1})}
\newcommand{\fig}[1]{Fig.~\ref{#1}}
\newcommand{\figs}[1]{Figs.~\ref{#1}}

\def \TT   {{\cal T}}
\def \AA   {{\cal A}}

\DeclareMathOperator*{\argmax}{arg\,max}

\theoremstyle{plain}
\newtheorem{theorem}{Theorem}
\newtheorem{lemma}{Lemma}

\newtheorem{corollary}{Corollary}

\theoremstyle{definition}

\newtheorem{definition}{Definition}
\theoremstyle{remark}
\newtheorem{remark}{Remark}

\newtheorem{claim}{Claim}

\input{tcilatex}

\graphicspath{{img/}}

\begin{document}

\title{Sample Greedy Based Task Allocation for Multiple Robot Systems}
\author{Hyo-Sang Shin*, Teng Li and Pau Segui-Gasco\thanks{
Hyo-Sang Shin, Teng Li and Pau Segui-Gasco are with the School of Aerospace, Transport and Manufacturing, Cranfield University, Cranfield MK43 0AL, UK
 (email: $\left\{\text{h.shin,tengli}\right\}$@cranfield.ac.uk)}
\thanks{ *Corresponding author.}
}
\markboth{Journal of \LaTeX\ Class Files}%
{Shell \MakeLowercase{\textit{et al.}}: Bare Demo of IEEEtran.cls for Journals}
\maketitle

\begin{abstract}
This paper addresses the task allocation problem for multi-robot systems. The main issue with the task allocation problem is inherent complexity that makes finding an optimal solution within a reasonable time almost impossible. To hand the issue, this paper develops a task allocation algorithm that can be decentralised by leveraging the submodularity concepts and sampling process. The theoretical analysis reveals that the proposed algorithm can provide approximation guarantee of $1/2$ for the monotone submodular case and $1/4$ for the non-monotone submodular case in average sense with polynomial time complexity. To examine the performance of the proposed algorithm and validate the theoretical analysis results, we design a task allocation problem and perform numerical simulations. The simulation results confirm that the proposed algorithm achieves solution quality, which is comparable to a state-of-the-art algorithm in the monotone case, and much better quality in the non-monotone case with significantly less computational complexity.
\end{abstract}



\section{Introduction}

\gls{mrs} have been gaining increasing attention thanks to their ability to coordinate simultaneous or co-operate to achieve common goals. \gls{mrs} provides some fundamental strengths that could not be achieved with single agent systems, e.g., increased flexibility, enhanced reliability and resilience, simultaneous broad area coverage or capability to operate outside the communication range of base stations. Specific applications under consideration include surveillance and monitoring \cite{Cole2006a,Pitre12,Lim2016,Li2017403}, border patrol \cite{Kingston_08,Bein2015479,Minaeian20161005}, police law enforcement \cite{Puri04,Carapezza_99}, forest-fire localisation \cite{Merino2006a,Belbachir2016153} and precision agriculture\cite{Barrientos2011a}. 

Efficient cooperation of \gls{mrs} is a vital part for their successful operations and effective assignment, termed as task allocation, of the available resources is the key enabler of such cooperation. This is because the strength of \gls{mrs} hinges on the distributed nature of the sensing resources available, making the successful assignment of these resources key to maximise its operational advantages. 

The main issue with the task allocation problem is that it has been proven to be NP-hard in general \cite{shin2014}. This means that task allocation problems require exponential time to be solved optimally, thus require a careful craft of approximation strategies. A key trade-off that must be performed concerns the optimality of the solution versus the computational complexity. 

There have been extensive studies to handle the NP-hardness issue in task allocation. The representative task allocation algorithms can be generally classified into three categories: exact approach \cite{HUMO1970122,bertsekas1999nonlinear,boggs2000sequential}, heuristic approach \cite{elsayed2014new,rathi2016improved,du2016ant} and approximation approach \cite{badanidiyuru2014fast,buchbinder2014submodular,sviridenko2017optimal}. Exact algorithms find optimal solutions, but they cannot normally guarantee polynomial time complexity. When the problem complexity becomes high or limited time is available for a solution, exact algorithms are often inapplicable. Heuristic algorithm can deliver good feasible solutions with certain convergence speed for the tested objective functions. As a consequence, most of the algorithms presented in this case are heuristic in their nature, see for example \cite{Dias2006} for a classic review on market based algorithms,  \cite{Yan2013a} and \cite{shin2014} for recent surveys. The issue with the heuristic approach is that they can guarantee neither optimality, nor the quality of the solution.  Approximation algorithms find an approximated solution that balances optimality and the computational complexity. Moreover, they typically provide mathematical guarantee on the quality of solution and computational complexity if the problem satisfies certain conditions, e.g., submodularity condition. Note that some approaches exploit advantages of two or more different types of approaches and hence those are called hybrid approach \cite{yuan2014hybrid}. 

Task allocation algorithms under consideration in this paper are the ones that exploit the approximation approach and also that can be leveraged for decentralised task allocation. Note that the centralised solution of the task allocation problem involves having to communicate all the agents and environment data to a centralised entity. This may not be possible in some realistic scenarios because relying on a central entity could remove resilience or the bandwidth to communicate all the information may not be available. Capability to be decentralised can relax such issues and thus enabling decentralisation provides an option that could be beneficial for extending \gls{mrs} operations in practice.  

Most of the decentralised task algorithms have taken two qualitatively different approaches. The first approach targets a trivial instance by assuming linearity of the utility function, enabling optimal algorithms \cite{Bertsekas1991, Liu2013b, Choi, ismail2017decentralized, Moon2012a}. The second approach is the heuristic approach \cite{parker1998alliance, dias2004traderbots, gerkey2002sold}. A great breakthrough was the introduction of a new distinct approach: an algorithm that used non-trivial properties on the objective function to enable decentralised constant-factor approximation algorithms. In \cite{Choi}, Choi et al presented the Consensus-Based Bundle Auction (CBBA) algorithm, which was the first decentralised approximate algorithm that provided a solution guaranteed to be within a constant factor of the optimal. By assuming that the utility functions of the agents are monotone non-decreasing and submodular, whose definitions will be given in Section \ref{sec:1}, their algorithm is proven to provide a solution with guarantee of at least 50\% of the value of the optimal solution.

There are deep theoretical reasons behind the choice of submodularity, and they are intimately connected with the tractability of the task allocation problem. Note that submodularity means that the marginal value that an agent obtains by executing an extra task diminishes as the tasks that need to be carried out by the agent increases. The formal definition and its meaning will be discussed in detail in Section \ref{sec:2}. The task allocation problem can be viewed as optimisation of a set function subject to a matroid constraint. Hence, submodularity plays a key role defining the tractability boundaries of the task allocation problem. The matroid constrained optimisation problem is  generally NP-Hard and inapproximable \cite{svitkina2011submodular}. Although it remains NP-hard, the matroid constrained maximisation problem becomes approximable if submodularity is assumed. In the late 1970s, a seminal paper by Nemhauser et. al. \cite{Nemhauser1978} proved that the greedy algorithm achieves a ${1}/{2}$ approximation guarantee with monotone submodular functions. This factor lies at the core of the reason why CBBA provides a constant factor approximation with monotone submodular functions as it is essentially a decentralised implementation of greedy algorithm. 

Immediate research questions would be whether ${1}/{2}$ is the best constant factor approximation that can be achieved for the task allocation problem with monotone submodular functions and what about non-monotone submodular functions. Answering these two questions, Segui-Gasco and Shin developed a decentralised task allocation algorithm that can achieve $(1-1/e-\epsilon)$ approximation guarantee for monotone submodular functions and $(1/e-\epsilon)$ for non-monotone submodular functions \cite{segui2017fast}. The  proposed algorithm leveraged a recent breakthrough in submodular optimisation, that is called measured continuous greedy algorithm. This algorithm, which is originally proposed by Feldman et al \cite{Feldman2011}, obtains $(1-{1}{/e})$ approximation for matroid-constrained monotone submodular maximisation and a ${1}{/e}$ factor for the non-monotone case. The issue is that the measured greedy algorithm might be computationally too complex to be directly utilised for task allocation. Therefore, Segui-Gasco and Shin  \cite{segui2017fast} developed a smoother version of measured continuous greedy and mitigated the computational complexity issue. Although the smoother version relaxes the computational complexity in orders of magnitude to the original measured continuous greedy, the complexity is still high compared with CBBA and it might be inefficient to be used for large scale \gls{mrs}.

This paper aims to develop a decentralised task allocation algorithm that can be implemented in practice for large scale \gls{mrs} based on submodular maximisation. The main objective is to provide mathematical guarantees on optimality not only for monotone submodular functions, but also for non-monotone ones with light computational complexity. To achieve the main objective, this paper proposes to leverage a sampling process and the main concepts of CBBA. The proposed algorithm is named as  DSTA (Decentralised Sample-based Task Allocation). In DSTA, each agent first randomly selects tasks from the ground task set with a probability of $p \in (0, 0.5]$. Then, each agent selects the task that provides the maximum marginal gain only from the task samples, not from all the tasks. Next, all agents agree on one optimal task-agent pair through maximum consensus protocol. The greedy selection process repeats until it reaches certain stop criteria. The proposed DSTA  algorithm unlocks three types of advantages. First, the algorithm achieves approximation guarantees in expectation for both monotone and non-monotone submodular functions. Second, the computational complexity can be relaxed as the algorithm considers only sampled tasks, not all tasks in the greedy selection phase. Third, DSTA enables the trade-off between the approximation guarantee and computational complexity by adjusting the sampling probability $p$.

The advantages of the proposed algorithm are examined mainly by theoretical analysis. To demonstrate and confirm the analysis results, this paper defines a simple benchmark mission scenario and performs numerical simulations. The benchmark scenario selected is a multi-UAV surveillance mission and two different types of submodular functions to be maximised are defined in the scenario. From the simulation results, performance, i.e. optimality and computational complexity, of the proposed algorithm is thoroughly investigated and compared with that of CBBA.

The rest of paper is organised as follows.  Section \ref{sec:1} presents some preliminaries and backgrounds which will be essential for algorithm development and analysis. Section \ref{sec:2} develops a new task allocation algorithm, namely DSTA algorithm, and its analysis is detailed in Section \ref{sec:3}. The performance and validity of the analysis results are investigated through numerical simulations in Section \ref{sec:4}. Section \ref{sec:5} offers some conclusions and discusses future research directions.

\section{Preliminaries}
\label{sec:1}

This section provides some necessary preliminaries for the development and analysis of the proposed decentralised task allocation algorithm.

Following general definition of the task allocation problem from \cite{Dias2006}, the particular instance of the task allocation problem under consideration in this paper is defined as:

\begin{definition}[Task allocation] 
\label{def:tap}
Given a set of tasks $\mathcal{T}$, a set of agents $\mathcal{A}$, and a non-negative submodular function for each agent $a\in\mathcal{A}$ specifying the utility of completing each subset of tasks $f_a:2^{\mathcal{T}} \rightarrow \mathbb{R}^{+}$, find a non-overlapping allocation, $\mathcal{S}^*\in \mathcal{A}^\mathcal{T}$, that maximises a global objective function $\mathcal{F}:\mathcal{A}^\mathcal{T}\rightarrow\mathbb{R}^+$ defined as $\mathcal{F}(S)=\sum_{a\in\mathcal{A}}{f_a({S_a})}.$
\end{definition}

The utility for each agent $a$ is submodular of which definition given in the following.

\begin{definition}[Submodularity \cite{krause2014submodular}]
\label{def:submodularity}
Let $\mathcal{N}$ be a finite set. A real-valued set function $f:2^{\mathcal{N}}\rightarrow\mathbb{R}$ is \textit{submodular} if, for all $X,Y\subseteq\mathcal{N}$,
\begin{equation}
f(X)+f(Y)\geq f(X\cap Y)+f(X\cup Y).
\end{equation}
Equivalently, for all $A\subseteq B \subseteq \mathcal{N}$ and $u\in\mathcal{N}\backslash B$,
\begin{equation}
\label{def:diminishing returns}
f(A\cup\{u\})-f(A)\geq f(B\cup\{u\})-f(B).
\end{equation}
\end{definition}

Note that the submodular function considered in this paper is normalised (i.e. $f(\emptyset)=0$)  and non-negative (i.e. $f(S) \geq 0$ for all $S \subseteq \mathcal{N}$). This is because the submodular maximisation problem is inapproximable if the submodular function can take negative values \cite{feige2011maximizing}. 

Submodularity is quite an intuitive notion. It simply requires that the marginal value that an agent obtains by executing an extra task diminishes as the tasks that ought to be carried out by the agent increases. Many engineering problems such as control and estimation problems are not obviously convex; but scientists and engineers have devised useful models to ``convexify" the problem, to enable practical algorithms that result in immensely useful applications. Like convexity, we believe that submodularity is a good modelling tool in designing utility functions for the task allocation problem. In our opinion, submodularity could be seen from the same perspective: a useful model that can be leveraged to induce the desired behaviour in \gls{mrs}. In the machine learning community, there has been a great effort spearheading this idea: finding suitable submodular models to solve inherently discrete tasks, such as summarising documents, scene segmentation, or pattern discovery \cite{song2014weakly, mirzasoleiman2016fast, Bilmes2017}. Task allocation is an inherently discrete problem and thus ``submodularising"  task allocation problems should yield useful applications of efficient  \gls{mrs} cooperation.

\eqn{def:diminishing returns} is known as the diminishing returns which is one of core properties of submodular functions. The diminishing return means that  the marginal gain of a given element $u$ will never increase as more elements have already been added into the set $S$. The range of applications holding this property could be wide \cite{Choi}, e.g. a surveillance mission in which the time that an agent will be able to monitor an additional point decreases as a given agent is assigned more points to monitor. of The definition of marginal gain is provided in Definition \ref{def:marginal_gain}.

\begin{definition}[Marginal gain \cite{krause2014submodular}]
\label{def:marginal_gain}
For a set function $f:2^{\mathcal{N}}\rightarrow\mathbb{R}$, $S \subseteq \mathcal{N}$ and $u \in \mathcal{N}$, the \textit{marginal gain} of $f$ at $S$ with respect to $u$ is defined as
\begin{equation}
\Delta f(u|S):=f(S \cup \{u\})-f(S).
\end{equation}
\end{definition}

As discussed in introduction, developing an algorithm that works with non-monotone submodular utilities could be of significant importance since non-monotonicity is a feature that arises naturally in many practical scenarios. For example, in a multi-robot surveillance mission, if a robot is assigned too many targets to track it is possible that it might end up spending its time travelling between targets and not gathering enough useful information at the targets' locations. Therefore, adding tasks to a robot's assignment could, indeed, reduce the utility obtained. Monotone submodular functions are structurally ill-suited to model such a scenario since, by definition, they do not contemplate a reduction in value due to an excessively large number of tasks. Therefore, this paper considers the submodular function that could be either monotone and non-monotone as described in Definition \ref{def:tap}. 

\begin{definition}[Monotonicity \cite{krause2014submodular}]
\label{def:monotonicity}
A set function $f:2^{\mathcal{N}}\rightarrow\mathbb{R}$ is \textit{monotone} if, for every $A \subseteq B \subseteq \mathcal{N}$, $f(A) \leq f(B)$. And $f$ is \textit{non-monotone} if it is not monotone.
\end{definition}

In Definition \ref{def:tap}, the non-overlapping allocation implies that one task can be allocated at most to one agent, but one agent can take more than one tasks. Hence, the task allocation problem described in Definition \ref{def:tap} can be reduced to submodular maximisation subject to a partition matroid constraint. The definition of matroid is given in Definition \ref{def:matroid}. 

\begin{definition}[Matroid \cite{badanidiyuru2014fast}]
\label{def:matroid}
A matroid is a pair $\mathcal{M}=(\mathcal{N},\mathcal{I})$ where $\mathcal{N}$ is a finite set and $\mathcal{I} \subseteq 2^\mathcal{N}$ is a collection of independent sets, satisfying:
\begin{itemize}
\item $\emptyset \in \mathcal{I}$ 
\item $A \subseteq B, B \in \mathcal{I} ~ \Rightarrow~ A \in \mathcal{I}$ 
\item $A, B \in \mathcal{I},|A|<|B| ~ \Rightarrow ~ \exists~b \in B \backslash A ~\mbox{such that}  ~A \cup \{b\} \in \mathcal{I}$
\end{itemize}
\end{definition}

Representative matroid constraints are uniform matroid and  partition matroid constraints. In the uniform matroid constraint, which is also called cardinality constraint, any subset $S \subseteq \mathcal{N}$ satisfying $|S|\leq k$ is independent. The partition matroid constraint means that a subset $S$ can contain at most one element from each partition. Therefore, in our task allocation problem, each partition is a set of agent-task pairs for each task. 

The following is an important claim \cite{buchbinder2014submodular} that provides the mathematical foundation for the analysis of the proposed task allocation algorithm, especially for the non-monotone submodular case.

\begin{claim}
\label{thm:main_claim}
Let $h:2^\mathcal{N}\rightarrow\mathbb{R}_{\geq0}$ be non-negative and submodular, and let $S$ be a random subset of $\mathcal{N}$ where each element appears with probability at most $p$ (not necessarily independently). Then, $\mathbb{E}[h(S)]\geq(1-p)h(\emptyset)$.
\end{claim}

\section{Algorithm}
\label{sec:2}

This section develops a decentralised task allocation algorithm and provides an equivalent centralised version that will be useful for the theoretical analysis.

\begin{algorithm}
\caption{DSTA for Agent $a$}
\label{alg:dsta}
\textbf{Input:} $f_a:2^\mathcal{T}\rightarrow\mathbb{R}_{\geq0}, \mathcal{T}, p$   \\
\textbf{Output:} A set $\mathcal{T}_a\subseteq\mathcal{T}$
\begin{multicols}{2}
\begin{algorithmic}[1]
\State $\mathcal{N}_a \leftarrow \emptyset$, $\mathcal{T}_a \leftarrow \emptyset$
\For {$j \in \mathcal{T}$}
	\State with probability $p$, 
	\State $\mathcal{N}_a \leftarrow \mathcal{N}_a \cup \{j\}$
\EndFor 
\While {$\exists ~ j \in \mathcal{N}_a$ such that $\Delta f_a(j| \mathcal{T}_a)>0$}
	\State $j_a^*\leftarrow \argmax\limits_{j\in \mathcal{N}_a}f_a(j| \mathcal{T}_a)$
	\State $\omega_a^* \leftarrow \Delta f_a(j_a^*| \mathcal{T}_a)$
	\State $a^*, j_{a^*}^* \leftarrow MaxCons(a,j_a^*,\omega_a^*)$
	\If {$a^*==a$}
		\State $\mathcal{T}_a \leftarrow \mathcal{T}_a \cup \{j_a^*\}$
		\State $\mathcal{N}_a \leftarrow \mathcal{N}_a \backslash \{j_a^*\}$
	\Else
		\If {$ j_{a^*}^* \in \mathcal{N}_a$}
			\State $\mathcal{N}_a \leftarrow \mathcal{N}_a \backslash \{j_{a^*}^*\}$
		\EndIf
	\EndIf
\EndWhile\\
\Return {$\mathcal{T}_a$}
\end{algorithmic}
\end{multicols}
\end{algorithm}

The proposed DSTA algorithm is summarised in Algorithm \ref{alg:dsta}. The main proposition in this paper is inclusion of a sampling process at the beginning the algorithm and integrate with the main concepts of CBBA. Therefore, in the proposed algorithm, each agent first randomly samples tasks from the ground task set with a probability $p$ to form its own set of task samples, $\mathcal{N}_a$ (lines 2 -- 5). Then, each agent selects a task that provides the maximum marginal gain from its own task samples $\mathcal{N}_a$ using the auction process (line 7) and saves the corresponding marginal gain (line 8).  Once each agent selects a task at each step, they communicate agent id, the selected task and the corresponding marginal gain with their neighbour agents. Then, they negotiate and agree on one task-agent pair that provides the largest marginal gain through maximum consensus protocol. $MaxCons(a,j_a^*,\omega_a^*)$ in the line 9 of Algorithm \ref{alg:dsta} is the max-consensus function, which represents the negotiation among all agents. For the max-consensus, each agent $a$ sends its current best marginal value $\omega_a^*$ together with corresponding agent id $a$ and task id $j_a^*$ to neighbour agents. At the same time, each agent receives the same information from all other neighbour agents. After agreeing through max-consensus, the agent of the best agent-task pair adds the corresponding task to its task bundle and remove the task from the task samples (lines 10 -- 12). If any agent includes the task of the best agent-task pair in the task samples, they remove that task from their task samples (lines 14 -- 16). Each agent repeats these procedures until there is no task to be allocated. 

\begin{remark}
Thanks to sampling, each agent in DSTA is required to evaluate the function values only for a portion of entire tasks, i.e. only for sampled tasks. This should enable to accelerate task allocation by some degree, which will depend on the sampling probability $p$. 
\end{remark}

Note that each agent in CBBA first constructs a bundle of tasks such that it optimises the score function, by continuously adding tasks until the algorithm is no longer able to add any other task. The score function considered in CBBA is a monotone submodular function that is a function of the task bundle and the path \cite{Choi} . Then, CBBA applies a consensus strategy to enable convergence on the list of winding bids and agents. For the consensus strategy, a set of sophisticated decision rules was proposed and the convergence characteristics were analysed. The consensus strategy in \cite{Choi} allows conflict resolution over all tasks and does not limit the network to a specific structure.  

Different from CBBA, our proposed algorithm utilises a simple strategy to build a bundle of tasks for each agent: each agent finds one best task and then applies a simple consensus strategy on the best tasks at each step. Following this strategy, all agents agree on one best task-agent pair and only the agent in the best task-agent pair adds the corresponding task to its bundle. By repeating the process until no agent can add anymore task to its bundle, each agent is able to construct individual bundles that are conflict-free. If necessary, the proposed algorithm can also use the same consensus strategy as CBBA.  

\begin{remark}
Once the bundles are constructed, CBBA initiates the consensus procedure and it might be able to reduce communication burden. However, the consensus strategy in CBBA could require a high number of function evaluations. Note that if a bundle of an agent includes any of tasks updated during  consensus iterations in CBBA, the agent must release all of the tasks that were added to the bundle after the updated task and reconstruct the bundle at each iteration. Reconstructing the bundle requires function evaluations. On the other hand, our strategy and requires the same number of consensus procedures as the number of tasks. However, the strategy is relatively simple and doesn't require the agents to reconstruct their task bundles. Consequently, the number of function evaluations in our proposed strategy could be reduced, compared with the consensus strategy in CBBA.  
\end{remark}

\begin{remark}
Developing a max-consensus protocol is beyond the scope of this study. Hence, a simple max-consensus protocol is utilised for case studies in Section \ref{sec:4}. There are many max-consensus algorithms available in existing literature. Also, the convergence characteristics of such algorithms are well studied, considering various practical aspects of communication, e.g. dynamic network, asynchronous communication, time delay, etc. \cite{cortes2008distributed,giannini2016asynchronous,iutzeler2012analysis,olfati2004consensus}. For the application of the proposed DSTA algorithm, one can select an efficient consensus protocol suitable for the communication network placed. 
\end{remark}

\begin{algorithm}
\caption{A Centralised Version of DSTA}
\label{alg:dsta_cv}
\textbf{Input:} $f_a:2^\mathcal{T}\rightarrow\mathbb{R}_{\geq0}~ \forall a\in \mathcal{A}, \mathcal{T}, p$   \\
\textbf{Output:}  Sets $\mathcal{T}_a\subseteq\mathcal{T}~\forall a\in \mathcal{A}$

\begin{multicols}{2}
\begin{algorithmic}[1]
\For {$a \in \mathcal{A}$}
	\State $\mathcal{N}_a \leftarrow \emptyset$, $\mathcal{T}_a \leftarrow \emptyset$
	\For {$j \in \mathcal{T}$}
		\State with probability $p$, 
		\State $\mathcal{N}_a \leftarrow \mathcal{N}_a \cup \{j\}$
	\EndFor 
\EndFor
\While {$\exists~ a\in \mathcal{A}$ and $j \in \mathcal{N}_a$ such that $\Delta f_a(j| \mathcal{T}_a)>0$}
	\For {$a\in \mathcal{A}$}
		\If {$\exists~j \in \mathcal{N}_a$ s.t. $\Delta f_a(j| \mathcal{T}_a)>0$}
			\State $j_a^*\leftarrow \argmax\limits_{j\in \mathcal{N}_a}f_a(j| \mathcal{T}_a)$
			\State $\omega_a^* \leftarrow \Delta f_a(j_a^*| \mathcal{T}_a)$
		\EndIf
	\EndFor
	\State $a^*, j_{a^*}^* \leftarrow \argmax \limits_{a\in \mathcal{A}, j_a^* \in \mathcal{N}_a} \omega_a^*(a,j_a^*)$
	\For{$a\in \mathcal{A}$}
		\If {$a^*==a$}
			\State $\mathcal{T}_a \leftarrow \mathcal{T}_a \cup \{j_a^*\}$
			\State $\mathcal{N}_a \leftarrow \mathcal{N}_a \backslash \{j_a^*\}$
		\Else
			\If {$ j_{a^*}^* \in \mathcal{N}_a$}
				\State $\mathcal{N}_a \leftarrow \mathcal{N}_a \backslash \{j_{a^*}^*\}$
			\EndIf
		\EndIf
	\EndFor
\EndWhile\\
\Return {$\mathcal{T}_a ~ \forall a\in \mathcal{A}$}
\end{algorithmic}
\end{multicols}
\end{algorithm}

Under the convergence of the max-consensus, the decentralised algorithm can be understood in an equivalent centralised version for the ease of analysis. Algorithm \ref{alg:dsta_cv} summarises this equivalent centralised version of DSTA. Note that Algorithm \ref{alg:dsta} is for each agent, whereas Algorithm \ref{alg:dsta_cv} is for all the agents since it is the centralised version.

\section{Analysis}
\label{sec:3}

This section analyses the performance, especially optimality and computational complexity, of the DSTA algorithm. Assuming convergence of the max-consensus, which is well studied in many existing literature \cite{cortes2008distributed,giannini2016asynchronous,iutzeler2012analysis,olfati2004consensus}, we perform analysis on the centralised version of DSTA. 

\subsection{Algorithm Analysis}
Consider the ground set as a set of task-agent pairs ($\mathcal{N}:= \mathcal{T} \times \mathcal{A}$) and each task-agent pair as an element of the ground set ($u_{j,a}:=j \times a ~ \forall j \in \mathcal{T}, a \in \mathcal{A}$). Then, the task-agent pairs can be considered as set elements and the task allocation problem can be understood as a submodular maximisation problem subject to a partition matroid constraint. Consequently, Algorithm \ref{alg:dsta_cv} can be further simplified to Algorithm \ref{alg:sample_greedy}. The resultant algorithm given in in Algorithm \ref{alg:sample_greedy} is equivalent to the sampling greedy algorithm for submodular maximisation subject to a partition matroid constraint. 

\begin{algorithm}
\caption{Sample Greedy}
\label{alg:sample_greedy}
\textbf{Input:} $f:2^\mathcal{N}\rightarrow\mathbb{R}_{\geq0}, \mathcal{N}, \mathcal{I}, p$   \\
\textbf{Output:} A set $S\subseteq\mathcal{I}$
\begin{multicols}{2}
\begin{algorithmic}[1]
\State $\mathcal{N}_s \leftarrow \emptyset, S \leftarrow \emptyset$
\For {$u \in \mathcal{N}$}
	\State with probability $p$, 
	\State $\mathcal{N}_s \leftarrow \mathcal{N}_s \cup \{u\}$
\EndFor 
\While {$\exists~u \in \mathcal{N}_s$ such that $S\cup \{u\} \in \mathcal{I}$ and $\Delta f(u|S)>0$}
	\State $u^*\leftarrow \argmax\limits_{u\in \mathcal{N}_s \backslash S}f(u|S)$
	\State $S \leftarrow S \cup \{u^*\}$
\EndWhile\\
\Return {$S$}
\end{algorithmic}
\end{multicols}
\end{algorithm}

This paper considers Algorithm \ref{alg:sample_greedy} as the baseline algorithm for the performance analysis. Note that \cite{feldman2017greed} provided a good analysis scheme for the sample greedy for  $k$-extendable systems. Since the partition matroid constraint is a special case of $k$-extendable systems, the analysis scheme in \cite{feldman2017greed} can be leveraged to prove the proposed DSTA algorithm. Therefore, we follow the analysis scheme in \cite{feldman2017greed}, but make necessary modifications to facilitate the partition matroid constraint. To ease the analysis, we can transform Algorithm \ref{alg:sample_greedy} into an equivalent version, which is summarised in Algorithm \ref{alg:e-sample_greedy}.

\begin{algorithm}
\caption{Equivalent Sample Greedy}
\label{alg:e-sample_greedy}
\textbf{Input:} $f:2^\mathcal{N}\rightarrow\mathbb{R}_{\geq0}, \mathcal{N}, \mathcal{I}, p$   \\
\textbf{Output:} A set $S\subseteq\mathcal{I}$
\begin{multicols}{2}
\begin{algorithmic}[1]
\State $S \leftarrow \emptyset, R \leftarrow \mathcal{N}, C \leftarrow \emptyset, Q \leftarrow OPT$
\While {$\exists~u \in R$ such that $S\cup \{u\} \in \mathcal{I}$ and $\Delta f(u|S)>0$}
	\State $c\leftarrow \argmax\limits_{u\in R}f(u|S)$
	\State $S_c\leftarrow S$
	\State $C \leftarrow C \cup \{c\}$
	\State \textbf{with} probability $p$ \textbf{do}
	\State ~~~~$S \leftarrow S \cup \{c\}, Q \leftarrow Q \cup \{c\}$
	\State ~~~~Let $K_c\subseteq Q \backslash S$ be the smallest set such that $Q \backslash K_c \in \mathcal{I}$
	\State \textbf{otherwise}
	\State ~~~~\textbf{if} {$c \in Q$} \textbf{then}
	\State ~~~~~~~~ $K_c \leftarrow \{c\}$
	\State ~~~~\textbf{else}
	\State ~~~~~~~~ $K_c \leftarrow \emptyset$
	\State ~~~~\textbf{end~if}
	\State $Q \leftarrow Q \backslash K_c$
	\State $R \leftarrow R \backslash \{c\}$
\EndWhile\\
\Return {$S$}
\end{algorithmic}
\end{multicols}
\end{algorithm}

Algorithm \ref{alg:e-sample_greedy} introduces four variables $C$, $S_c$, $Q$ and $K_c$ just for the convenience of analysis. It is clear that theses variables have no effect on the final output $S$ and thus Algorithm \ref{alg:e-sample_greedy} is equivalent to Algorithm \ref{alg:sample_greedy}. 

The meanings and roles of the four variables are given as follows. The variable $C$ is the set that contains all task-agent pairs that have already been considered by Algorithm \ref{alg:e-sample_greedy} no matter whether they are added to $S$ or not. 

$S_c$ is the set that contains the selected task-agent pairs at the beginning of the current iteration. At the end of the current iteration, $S_c = S \backslash \{c\}$ if $c$ is added into $S$ and $Q$, otherwise $S_c = S$. 

$Q$ is the set that bridges the solution $S$ and the optimal solution $OPT$. $Q$ is initialised as $OPT$ at the beginning of the algorithm and updates over time. Every task-agent pair that is added into $S$ is also added into $Q$. At the same time, a set $K_c$ is removed from $Q$ at each iteration to keep the independence of $Q$ if a task-agent pair $c$ is added into $Q$. Notice that, if a task-agent pair $c$ is already in $Q$ and is considered but not added into $S$ at the current iteration, then this task-agent pair $c$ should be removed from $Q$. 

$K_c$ is the set that is introduced to keep $Q$ independent. According to the matroid properties, the  Algorithm \ref{alg:e-sample_greedy} is able to remove a set $K_c \subseteq Q\backslash S$ which contains at most one task-agent pair from $Q$ if a task-agent pair is added into the currently independent set $Q$.

\subsection{Performance Analysis}

The main performance characteristics of the proposed DSTA algorithm are summarised in Theorem \ref{thm:main_theorem}. 

\begin{theorem}
\label{thm:main_theorem}
Suppose the max-consensus in Algorithm \ref{alg:dsta} assures the convergence. Then, the DSTA algorithm achieves an expected approximation guarantee of $\frac{p}{p+P_{max}}$ for monotone submodular objective functions and of $\frac{p(1-p)}{p+P_{max}}$ for non-monotone submodular utility functions with an expected total computational complexity of $O(pnr)$ and individual complexity of $O(pr^2)$ for each agent, where $p$ is the sampling probability, $P_{max}=\max (p, 1-p)$, $r$ is the number of tasks, i.e. $r=|\mathcal{T}|$, and $n$ is the number of task-agent pairs i.e. $n=|\mathcal{T}|\times|\mathcal{A}|$. 
\end{theorem}


The computational complexity of DSTA can be easily proven. As shown in Algorithm \ref{alg:dsta_cv}, there are at most $r$ rounds of auctions. In each round, each agent requires function evaluations at most $pr$ times. Since there are $|\mathcal{A}|$ number of agents, the total number of value oracle calls in each auction is $O(pn)$. Therefore, the total time complexity is $O(pnr)$. For each agent, the individual time complexity is equivalent to the total complexity divided by the number of agents i.e. $O(pr^2)$.

Let us now investigate the approximation guarantee of DSTA through Algorithm \ref{alg:e-sample_greedy}. Since the task-agent pairs that have never been considered by Algorithm \ref{alg:e-sample_greedy} have no contribution to the approximation guarantee, we only need to focus on the task-agent pairs that are contained in $C$. To prove the approximation guarantee, we will first show the bound of expectation of $f(S)$, i.e. $\mathbb{E} \left[ f(S) \right]$, with respect to $\mathbb{E}[f(S \cup OPT)]$  in Lemma \ref{thm:lemma_3}. Lemmas \ref{thm:lemma_1} and \ref{thm:lemma_2} will be required to prove the  bound of  $\mathbb{E} \left[ f(S) \right]$ in Lemma \ref{thm:lemma_3}. 

\begin{lemma}
\label{thm:lemma_1}
$\mathbb{E}[|K_c \backslash S|] \leq P_{max}$ where $P_{max} = \max(p, 1-p)$.
\end{lemma}
\begin{IEEEproof}
For the proof, we have two cases to analyse, depending on whether the current task-agent pair $c$ is already in $Q$ at the beginning of the iteration in Algorithm \ref{alg:e-sample_greedy}.

    1. If $c \in Q$ at the beginning of the iteration, then $K_c=\emptyset$ if $c$ is added to $S$ and $K_c=\{c\}$ if $c$ is not added to $S$. As $c$ is added to $S$ with probability $p$, the law of total probability yields:
\begin{equation}
\mathbb{E}[|K_c \backslash S|] = p \cdot |\emptyset|+(1-p)|\{c\}|=1-p
\end{equation}

    2. If $c \notin Q$ at the beginning of the iteration, then $K_c$ contains at most one element if $c$ is added to $S$. This is because of the property of independent systems. If $c$ is not added to $S$ then $K_c=\emptyset$. Therefore, we have
\begin{equation}
\mathbb{E}[|K_c \backslash S|] \leq p \cdot 1+(1-p)|\emptyset|=p
\end{equation}

    In summary, $\mathbb{E}[|K_c \backslash S|] \leq \max(p, 1-p)$.    
\end{IEEEproof}

\begin{lemma}
\label{thm:lemma_2}
$\mathbb{E}[f(S)] = p \sum \limits_{c \in C}\Delta f(c|S_c)$ .
\end{lemma}

\begin{IEEEproof}
According to the order by which the task-agent pairs are added into $S$, $f(S)$ can be written as
\begin{equation}
f(S) = f(\emptyset) + \sum \limits_{c \in S}\Delta f(c|S_c) = \sum \limits_{c \in S}\Delta f(c|S_c)
\end{equation}
Note that the second equality is because $f$ is assumed to be normalised, i.e. $f(\emptyset) = 0$. Moreover, we have the same number of $S_c$ sets to the number of $c$ elements by Algorithm \ref{alg:e-sample_greedy}. In each iteration, the task-agent pair $c$ that is being considered is added to $S$ with probability $p$. If this task-agent pair $c$ is added to $S$, then its marginal value $\Delta f(c|S_c)$ will be added to the current function value $f(S_c)$. Otherwise, the contribution of $c$ is 0. Therefore,
\begin{equation}
f(S) = \sum \limits_{c \in C}[\Delta f(c|S_c)_{c \in S} + 0 \cdot \Delta f(c|S_c)_{c \in C\backslash S}]
\end{equation}
By the law of total probability, we have
\begin{equation}
\mathbb{E}[f(S)] = p \cdot \sum \limits_{c \in C}\Delta f(c|S_c) + (1-p) \cdot 0 = p \cdot \sum \limits_{c \in C}\Delta f(c|S_c)
\end{equation}    
\end{IEEEproof}

\begin{lemma}
\label{thm:lemma_3}
$\mathbb{E}[f(S)] \geq \frac{p}{p+P_{max}}  \mathbb{E}[f(S \cup OPT)]$.
\end{lemma}

\begin{IEEEproof}
From Algorithm \ref{alg:e-sample_greedy} and definition of $Q$, it is clear that the set $Q$ is independent, i.e. $Q\in \mathcal{I}$ and $S$ is a subset of $Q$ i.e. $S \subseteq Q$.  Therefore, from Definition \ref{def:matroid}, we have $S \cup \{q\} \in \mathcal{I} ~ \forall q \in Q \backslash S$. By the termination condition, at the termination of Algorithm \ref{alg:e-sample_greedy}, $\Delta f(q|S) \leq 0 ~ \forall q \in Q \backslash S$ . Hence,
\begin{equation}
\sum\limits_{q \in Q \backslash S}\Delta f(q|S) \leq 0
\end{equation}
Let $Q \backslash S = \{q_1, q_2, \cdots, q_{|Q \backslash S|}\}$, then
\begin{align*}
f(S) &= f(Q) - \sum \limits_{i=1}^{|Q \backslash S|}\Delta f(q_i|S \cup \{q_1, \cdots, q_{i-1}\}) &\\
& \geq f(Q) - \sum\limits_{i=1}^{|Q \backslash S|}\Delta f(q_i|S)  &\text{(submodularity)} \\
& \geq f(Q)
\end{align*}

If $c$ is being considered, it implies that the marginal value of $c$ is no less than any other element from $K_c \backslash S$ at that iteration, i.e.
\begin{equation} 
\label{c vs o}   
\Delta f(c|S_c) \geq \Delta f(q| S_c), \; \forall q \in K_c \backslash S
\end{equation}

Additionally, any task-agent pair can be removed from $Q$ at most once. In other words, the task-agent pair that is contained in $K_c$ at one iteration is always different from other iterations when $K_c$ is not empty. Therefore, the sets $\{K_c\}_{c \in C}$ and sets $\{K_c \backslash S\}_{c \in C}$ are disjoint. From the definition of $Q$, $Q$ can be expressed as
\begin{equation}
\label{definition of O}
Q = (OPT \backslash \cup_{c \in C}K_c) \cup S = (S \cup OPT) \backslash \cup_{c \in C}(K_c \backslash S)
\end{equation}
Denote $C$ as $\{c_1, c_2, \cdots, c_{|C|}\}$. Then, it holds that $S_{c_i} \subseteq S \subseteq (S \cup OPT) \backslash \cup_{c \in C}(K_c \backslash S)$. From \eqn{definition of O}, we obtain
\begin{align*}
f(Q) &= f((S \cup OPT) \backslash \cup_{c \in C}(K_c \backslash S)) &\\
&= f(S \cup OPT)- \Delta f(\cup_{c \in C}(K_c \backslash S)|(S \cup OPT)\backslash \cup_{c \in C} (K_c \backslash S)) &\\
&= f(S \cup OPT)- \sum \limits_{i=1}^{|C|} \Delta f((K_{c_i} \backslash S)|(S \cup OPT)\backslash \cup_{1\leq j \leq i} (K_{c_j} \backslash S)) &\\
& \geq  f(S \cup OPT) - \sum \limits_{i=1}^{|C|} \Delta f((K_{c_i} \backslash S)|S_{c_i})  &\text{(submodularity)} \\
& \geq  f(S \cup OPT) - \sum \limits_{i=1}^{|C|} \sum \limits_{q \in K_{c_i} \backslash S}\Delta f(q|S_{c_i})  &\text{(submodularity)} \\
& = f(S \cup OPT) - \sum \limits_{c \in C} \sum \limits_{q \in K_c \backslash S}\Delta f(q| S_c) &\\
& \geq f(S \cup OPT) - \sum \limits_{c \in C} \sum \limits_{q \in K_c \backslash S}\Delta f(c| S_c) &\text{(inequality (\ref{c vs o}))} \\
& = f(S \cup OPT) - \sum \limits_{c \in C}|K_c \backslash S|\Delta f(c| S_c) 
\end{align*}
    
    By taking expectation over $f(S)$, we have
\begin{align*}
\mathbb{E}[f(S)] &\geq \mathbb{E}[f(Q)] & \\
&\geq \mathbb{E}\left[ f(S \cup OPT) - \sum \limits_{c \in C}|K_c \backslash S|\Delta f(c|S_c) \right] &\\
&= \mathbb{E}[f(S \cup OPT)] - \mathbb{E}[|K_c \backslash S|] \cdot \sum\limits_{c \in C}\Delta f(c| S_c) &\\
&\geq \mathbb{E}[f(S \cup OPT)] - P_{max} \cdot \sum\limits_{c \in C}\Delta f(c| S_c) &\text{(Lemma \ref{thm:lemma_1})} \\
&= \mathbb{E}[f(S \cup OPT)] - P_{max} \cdot \frac{1}{p}\mathbb{E}[f(S)] &\text{(Lemma \ref{thm:lemma_2})}
\end{align*}
The result is clear by rearranging the above inequation.    
\end{IEEEproof}

    We are now ready to complete the proof of Theorem \ref{thm:main_theorem}.
\begin{IEEEproof}[Proof of Theorem \ref{thm:main_theorem}] In order to get the approximation guarantees for both monotone and non-monotone submodular utility functions, we need to analyze the relationship between $f(S \cup OPT)$ and $f(OPT)$ respectively. If $f$ is monotone, then 
\begin{equation}
\label{eqn:sub01}
f(S \cup OPT) \geq f(OPT)
\end{equation}
From Lemma \ref{thm:lemma_3} and \eqn{eqn:sub01}, it is clear that
\begin{equation}
\begin{split}
\mathbb{E}[f(S)] &\geq \frac{p}{p+P_{max}} \cdot \mathbb{E}[f(S \cup OPT)] \quad \text{(Lemma \ref{thm:lemma_3})} \\
&\geq  \frac{p}{p+P_{max}} \cdot \mathbb{E}\left[ f(OPT) \right]
\end{split}
\end{equation}
For the non-monotone case, define a new submodular and non-monotone function $h:2^\mathcal{N}\rightarrow\mathbb{R}_{\geq0}$ as $h(X)=f(X\cup OPT) ~\forall X \subseteq \mathcal{N}$. Since $S$ contains every element with probability at most $p$, Claim \ref{thm:main_claim} yields
\begin{equation}
\mathbb{E}[f(S \cup OPT)]=\mathbb{E}[h(S)] \geq (1-p)h(\emptyset)=(1-p)f(OPT)
\end{equation}
Hence,
\begin{equation}
\mathbb{E}[f(S)] \geq  \frac{p(1-p)}{p+P_{max}} \cdot \mathbb{E}\left[ f(OPT) \right]
\end{equation}
\end{IEEEproof}    

\begin{corollary} 
The trade-off between approximation guarantee and computational complexity can be controlled by adjusting the sampling probability $p$. When $p=1/2$, the DSTA algorithm achieves the best approximation guarantee of $\frac{1}{2}$ for the monotone case and of $\frac{1}{4}$ for the non-monotone case. 
\end{corollary}

\begin{IEEEproof}
Recalling that $P_{max}= \max(p, 1-p)$, we have
\begin{equation}
P_{max} = 
    \left\{ \begin{array}{lll} 
    1-p & \mbox{for} \; p \in (0, 0.5] \\
    p & \mbox{for} \; p \in (0.5, 1]
    \end{array} \right.
\end{equation}
Therefore, for $p \in (0, 0.5]$, the expected approximation ratios are obtained as:
\begin{equation}
\label{eqn:approx01}
\mathbb{E}[f(S)] \geq
     \left \{ \begin{array}{lll} 
     p \cdot f(OPT) & \text{if} \; f \; \mbox{is monotone} \\ 
     p(1-p) \cdot f(OPT) & \text{if} \; f \; \mbox{is non-monotone} 
     \end{array}\right.
\end{equation}
In case where $\; p \in (0.5, 1]$, it is trivial that
\begin{equation}
\label{eqn:approx02}
\mathbb{E}[f(S)] \geq 
    \left \{ \begin{array}{lll} 
    \frac{1}{2}f(OPT) & \mbox{if} \; f \; \mbox{is monotone}  \\ 
    \frac{1-p}{2}f(OPT) & \mbox{if} \; f \; \mbox{is non-monotone} 
    \end{array}\right.
\end{equation}
As shown in \eqns{eqn:approx01} and \eqnref{eqn:approx02}, there is no advantage in average sense to set the sampling probability bigger than $0.5$. For $p \in (0.5, 1]$, \eqn{eqn:approx02} clearly shows that the expected approximation ratio becomes stagnated in the monotone case and is monotonically decreasing for the non-monotone case as the sampling probability increases. Moreover, it is clear that the computational complexity increases as the sampling probability gets larger. On the other hand, for $p \in (0, 0.5]$, the sampling probability provides trade-off capability between the approximation ratio and computational complexity. As the probability increases for $p \in (0, 0.5]$, the expected approximation ratios improve for both monotone and non-monotone cases, but the computational complexity also increases. 
From \eqns{eqn:approx01} and \eqnref{eqn:approx02}, the best expected approximation guarantees can be readily obtained as:
\begin{equation}
\label{eqn:approx_best}
\mathbb{E}[f(S)] =
     \left \{ \begin{array}{lll} 
     \frac{1}{2} & \text{if} \; f \; \mbox{is monotone} \\ 
     \frac{1}{4} & \text{if} \; f \; \mbox{is non-monotone} 
     \end{array}\right.
\end{equation}
\end{IEEEproof}

\section{Numerical Simulations}
\label{sec:4}
This section verifies the proposed DSTA algorithm and compare its performance with that of a benchmark task allocation algorithm through numerical simulations. The benchmark algorithm selected is CBBA since it is one of most well known and widely applied decentralised task allocation algorithms. 

The application scenario considered is a multi-UAV surveillance mission. To demonstrate the applicability of the submodular maximisation based task allocation algorithms such as CBBA and DSTA, this paper formulates the multi-UAV surveillance problem as submodular maximisation problems subject to the matroid constraint. Hence, the utility function of each UAV is formulated as a submodular function. To examine the performance of the proposed DSTA in comparison with CBBA also in the non-monotone case, a value function is formulated, considering various different factors. 
    
In the simulations, it is assumed that there are 300 tasks to be carried out by different numbers of UAVs from 10 to 50 denoted as $N_a$. One task can be allocated to at most one agent, but one agent can carry out multiple tasks. The aim of the task allocation is to achieve the largest overall objective function value. The sampling probability used throughout the simulations is $0.5$.

\subsection{Monotone Case}
This subsection investigates the performance of the proposed DSTA algorithm in the monotone case. For fair comparison, this paper adopts a simple monotone value function utilised in \cite{Choi}. Suppose that, both agents and tasks are homogeneous. For each agent $a\in\mathcal{A}$, the value function is given as \cite{Choi}
 \begin{equation}
f_a(\mathcal{T}_a) = \sum\limits_{j \in \mathcal{T}_a} {\lambda _j^{\tau ({\mathbf{p}_a})}{b_j}}
\end{equation}
where $\mathcal{T}_a$ is the task set allocated to agent $a$, $\lambda_j<1$ is the discounting factor for each task $j\in\mathcal{T}$, $\tau({\mathbf{p}_a})$ is the estimated distance of the path $\mathbf{p}_a$, and $b_j$ is the static score associated with performing task $j$. In the simulations, $\lambda = 0.95$ and $b_j = 1$

\begin{figure}
\begin{subfigure}{.5\textwidth}
\centering
\includegraphics[width=\linewidth]{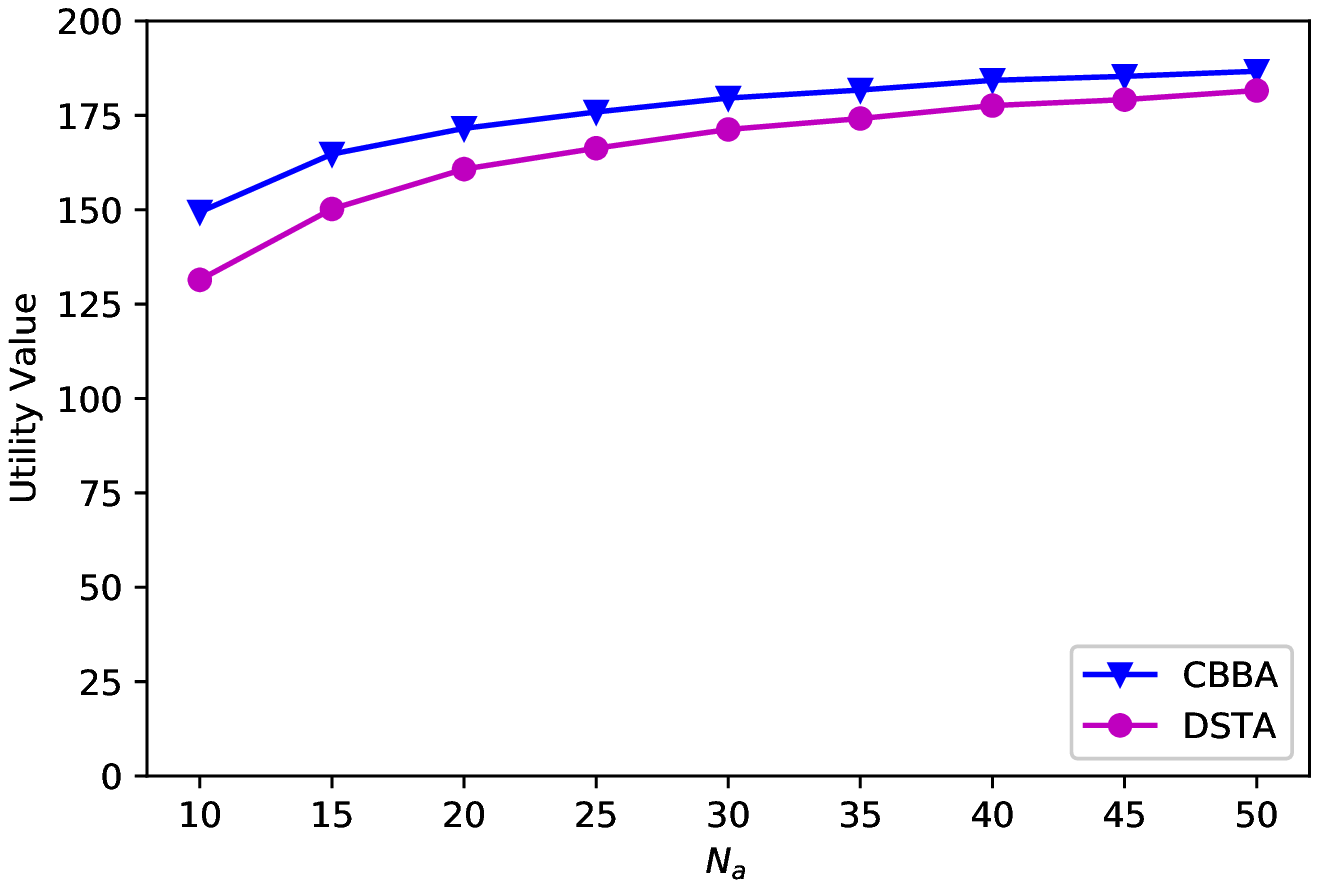}
\caption{Utility function}
\end{subfigure}
\begin{subfigure}{.5\textwidth}
\centering
\includegraphics[width=\linewidth]{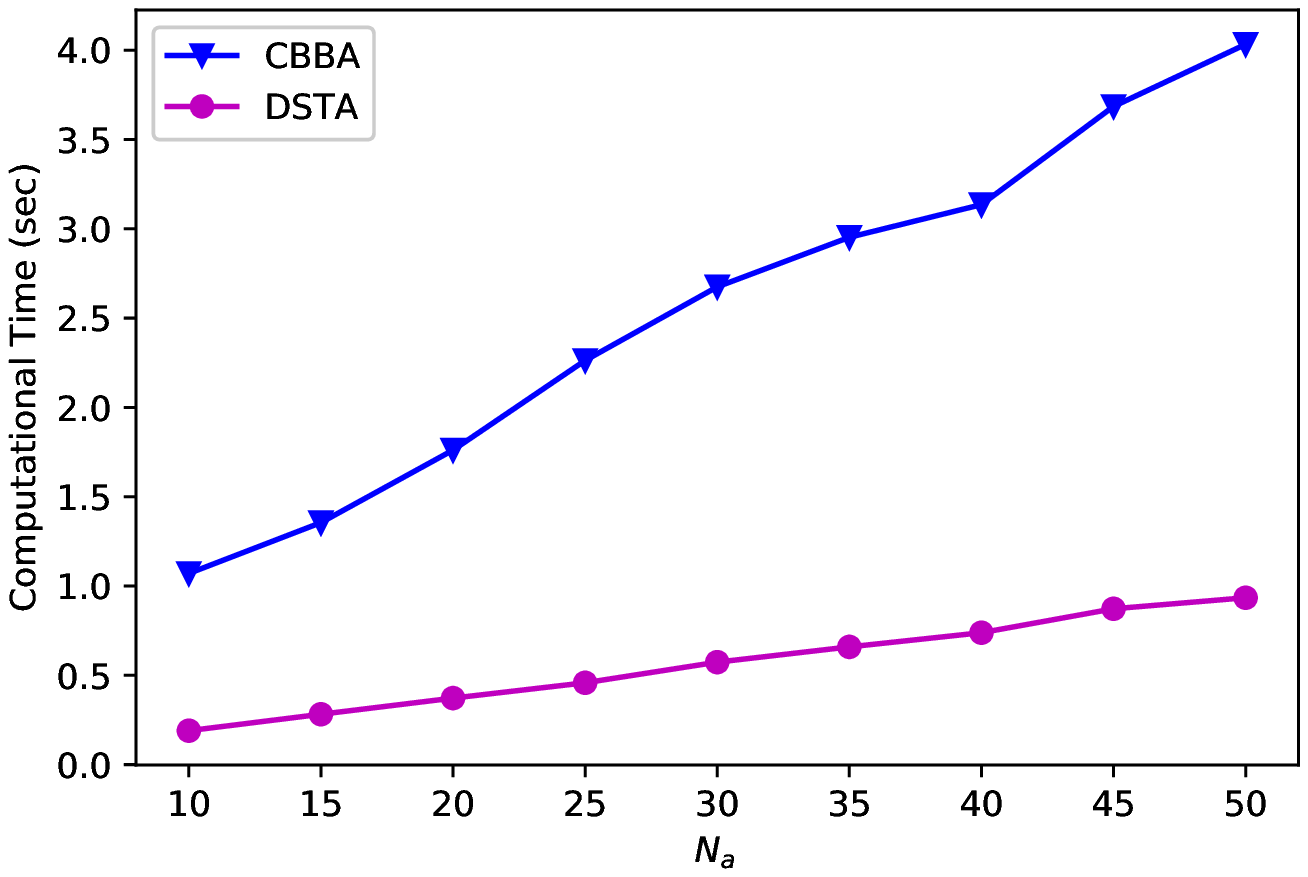}
\caption{Computational time}
\end{subfigure}
\caption{Performance comparison for the monotone submodular function ($N_t = 200$)}
\label{fig:monotone_all01}
\end{figure}

\begin{figure}
\begin{subfigure}{.5\textwidth}
\centering
\includegraphics[width=\linewidth]{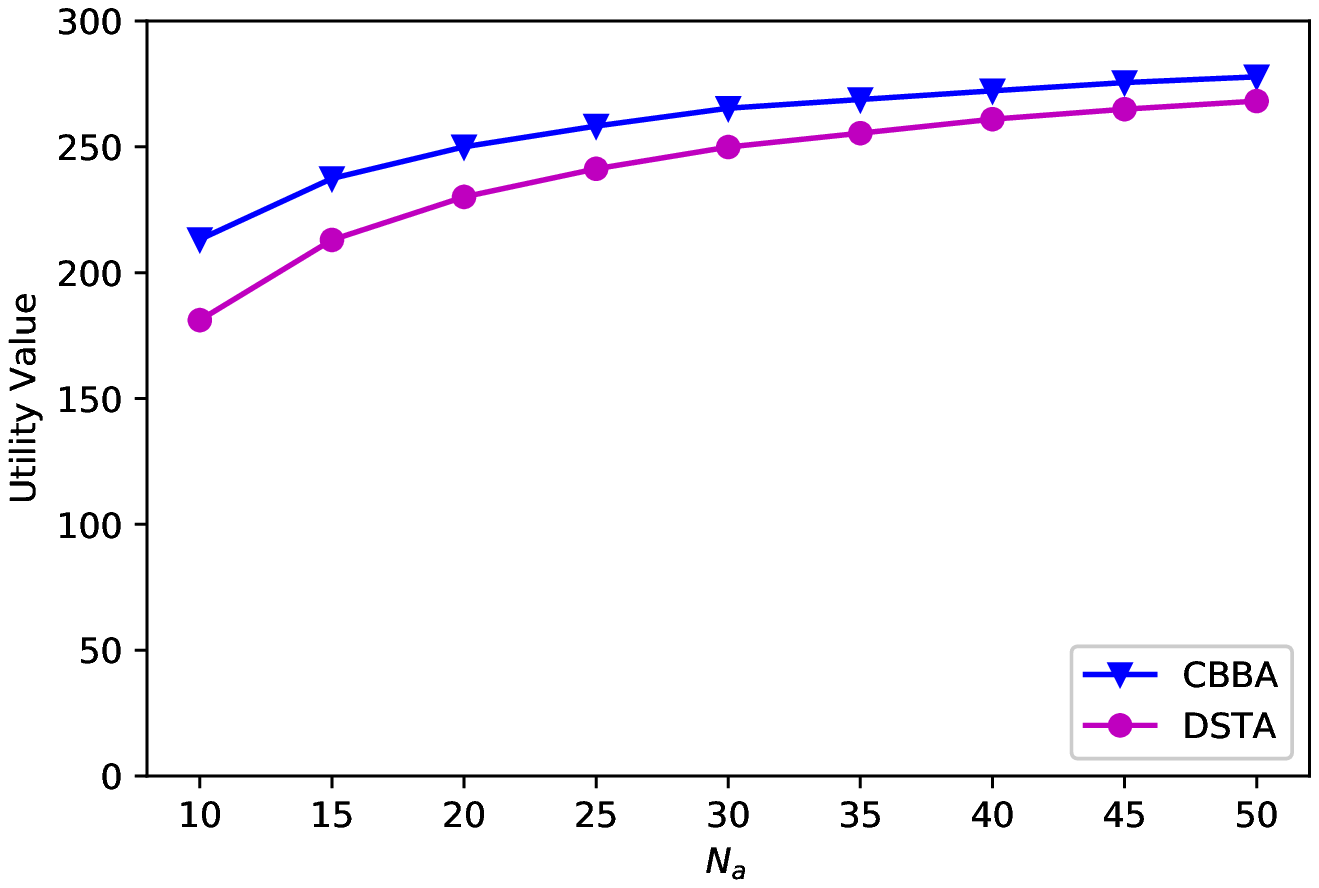}
\caption{Utility function}
\end{subfigure}
\begin{subfigure}{.5\textwidth}
\centering
\includegraphics[width=\linewidth]{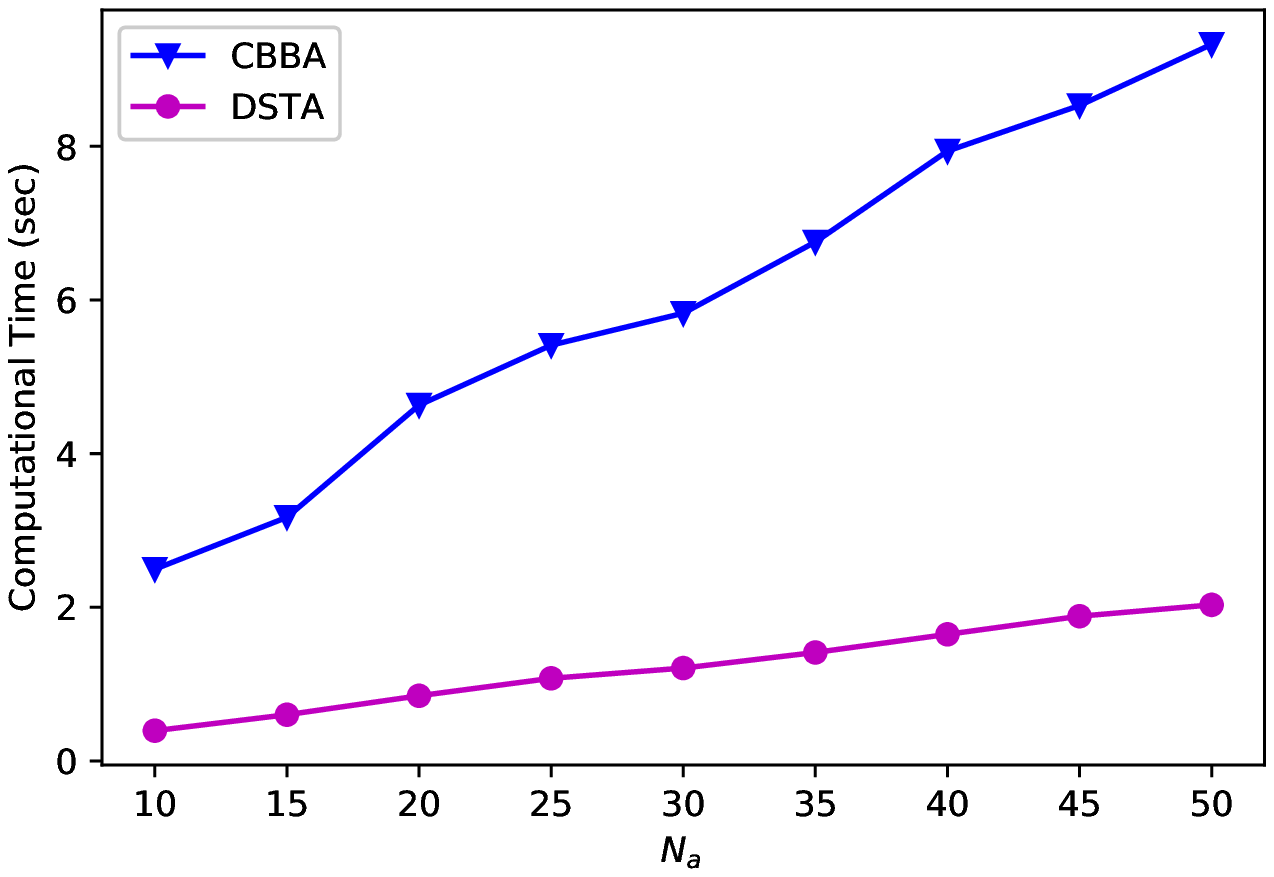}
\caption{Computational time}
\end{subfigure}
\caption{Performance comparison for the monotone submodular function ($N_t = 300$)}
\label{fig:monotone_all02}
\end{figure}

We run 10 Monte Carlo simulations in which agents and tasks are randomly placed on a $W \times W$ 2-D space ($W$ = 10 km). Depending on the number of tasks, there are two simulations cases: the number of tasks is 200 in the first case and 300 in the second case. The number of agents  incrementally increases from 10 to 50 in both cases. The simulation results are depicted in \figs{fig:monotone_all01} and \ref{fig:monotone_all02}. Note that the values in all figures in this subsection and the following subsection are the average values of the Monte Carlo simulation results. On the one hand, as shown in the figures, the proposed DSTA  algorithm achieves comparable quality of solutions, i.e. comparable values of the utility function, to CBBA , in both simulation cases. In average, the quality of DSTA solution is around 95\%, reaching to 97\%, in the first simulation case and is around 94\%, approaching to around 97\%, in the second case. On the other hand, computational time of the DSTA algorithm is significantly less compared with that of CBBA. Note that computational time is measured as the total time taken to complete task allocation for all the agents. The simulation codes are written in Python 3.3.1 and the Monte Carlo simulations were run in late 2014 Mac Mini with an Intel i7 3.0 GHz processor and 16Gb of RAM, running OSX 10.13.5, running both algorithms sequentially in a single core without any parallelism. It is worth noting that as the size of the problem increases, the difference between the two algorithms on the function value becomes tighter, but the difference on computational time becomes more significant. 


\subsection{Non-monotone Case}

For the non-monotone case, this paper adapts a surveillance mission in which a group of UAVs are deployed to explore an enemy area. The problem formulation and modelling are adapted in \cite{PauThesis}. The utility function adapted is non-monotone submodular and consists of various components. For the completeness, let us describe all the components of utility function and the resultant utility function.

\subsubsection{Values}

Each task could require different abilities or resources for its execution and might have different importance for the completion of mission. Considering these aspects, the value that an agent $a$ can obtain by performing task $j$ is formulated as
\begin{equation}
w_{aj} =: \sigma_j m_{aj}
\end{equation}
where $\sigma_j\in\mathbb{R}^{+}$ is the importance factor of task $j \in \TT$,  $m_{aj}\in\mathbb{R}_{\geq0}$ is the task-agent fitness factor.

For the simulations, this paper generates $m_{aj}$ for each task-agent pair in a randomly or arbitrary manner. However, the fitness match $m_{aj}$ can be defined arbitrarily as long as it is a positive real. 

By completing a set of tasks $\mathcal{T}_a\in\mathcal{T}$, the agent $a$ can obtain a value of:
\begin{equation}
\label{def:value}
v(\mathcal{T}_a) = \sum\limits_{j\in \mathcal{T}_a}w_{aj}
\end{equation}

\subsubsection{Probability of Mission Success}

The probability of being detected generally increases as an UAV is exposed longer in the enemy environment, i.e. as it carries out more tasks. Hence, the probability of being detected at the $(n+1)$th task, given that the UAV has executed previously $n$ tasks, is modelled as \cite{PauThesis}:
\begin{equation}
\label{eqn:pd_model}
P_{D,a} (n+1) = \Pr(\text{detection of $a$ at the $n+1$th task}|n) \triangleq \frac{P_{0,a}}{1-{\alpha_a n P_{0,a} }}
\end{equation}
where $P_{0,a} \in (0, 1)$ denotes the probability of UAV $a$ being detected by the enemy when it executes a single task without any previous tasks being executed. $\alpha_a\geq1$ is a parameter that governs how fast the detection probability increases. To guarantee $P_D ( |\TT| ) < 1$, $\alpha$ should be designed such that $\frac{P_0}{1-\alpha (|\TT|-1) P_0}<1$. 

Then, for the agent $a$ executing $n$ number of tasks, the probability of being detected can be quantified as:
\begin{equation}
P_{D,a}(n) = P_{D,a}(n-1) + \left(1-P_{D,a}(n-1) \right) \frac{P_{0,a}}{1-{\alpha_a (n-1) P_{0,a} }}
\end{equation}
where $P_{0,a} = P_{D,a}(0) $ by convention.

For simplicity, it is assumed that if an UAV has been detected, the mission has failed and thus produced a zero value \cite{PauThesis}. Then, we can define the probability of surviving,  $P_{S,a}: 2^\TT \rightarrow (0,1]$, during executing $\mathcal{T}_a$ task set as:
\begin{equation}
\label{def:probability}
P_{S,a}(\mathcal{T}_a) \triangleq 1- P_{D,a}(| \mathcal{T}_a |).
\end{equation}

\subsubsection{Inter-Task Effect}

As discussed, the probability of an UAV being detected increases as the UAV is exposed more to the adversarial mission environment. Therefore, it is risky to allocate too many important tasks to one single agent. To remove this fragility, inter-task penalties are introduced. 

Define $d^{a}_{ij}$ as the penalty of UAV $a$ when executing each pair $i,j \in \TT$ of distinct tasks in its task set. Then, the total penalty function, $g_a: 2^\TT \rightarrow \mathbb{R}^+$, for a given task set $\TT_a \subseteq \TT$ is defined as:
\begin{equation}
\label{def:effect}
g(\mathcal{T}_a) = \sum\limits_{i,j\in \mathcal{T}_a,i\neq j}d^{a}_{ij}
\end{equation}
To discourage concentration of important tasks in a single UAV, we set $d^{a}_{ij}= e^{\sigma_i\sigma_j}$ .

\subsubsection{Utility Function}
The utility function is obtained by combining all the terms described in \eqns{def:value}, \eqnref{def:probability} and \eqnref{def:effect}. That is, the value of a task set $\TT_a \subseteq \TT$ in agent $a\in \mathcal{A}$ is defined as the expected value obtained minus by inter-task penalties:
\begin{equation}
\label{eqn:utility_def}
f_a(\mathcal{T}_a)=\mathbb{E}_{S}[v_a(\mathcal{T}_a)]-\lambda_ag_a(\mathcal{T}_a)
\end{equation}
where $\lambda_a$ is a scaling factor ensuring that the value function is non-negative and $\mathbb{E}_{S}[\cdot]$ denotes the expected value that accounts the probability of the agent being detected and thus the corresponding value becoming zero. Plugging in all the definitions into \eqn{eqn:utility_def}, the final value function is given as:
\begin{equation}
\label{eqn:utility}
f_a(\mathcal{T}_a)=P_{S,a}(\mathcal{T}_a)\sum\limits_{j\in \mathcal{T}_a}w_{aj}-\lambda_a\sum\limits_{i,j\in \mathcal{T}_a, i\neq j}d^{a}_{ij}
\end{equation}
%


\subsubsection{Simulation Results}

Initial parameter setting is as follows: $\alpha_a=1, \lambda_a=0.01$ for all $a\in\mathcal{A}$. To guarantee $P_D ( |\TT| ) < 1$,  the initial detection probability is defined as $P_{0,a} = \frac{1}{1+\alpha (|\TT|)}$ for all $a \in \AA$.  

Assuming that there are certain tasks that are more important than other tasks, the importance factors of tasks are randomly drawn from two uniform distributions. The importance factors of $N_a$ number of tasks are drawn from a continuous uniform distribution between 5 and 7, i.e. $\mathcal{U}(5, 7)$. It is assumed that each of these $N_a$ tasks is suitable for one agent and hence we set the corresponding task-agent fitness factor, $w_{aj}$, as 0.3 while setting other factors related to each of these tasks as 0.1. The importance factors of rest of tasks are randomly drawn from $\mathcal{U}(0.5, 1.5)$, implying that they are less important than the $N_a$ tasks. All task-agent fitness factors corresponding to these tasks are randomly selected from a continuous uniform distribution between 0.1 and 1. 

First, let us compare the performance, especially the quality of solution and computational time, of the proposed DSTA algorithm with CBBA. Like for the monotone submodular function, we have two simulations cases depending on the number of tasks. For each case, we run 10 Monte Carlo simulations. The number of agents incrementally increases from 10 to 50 in both cases. The sampling probability $p$ is set to be equal to $1/2$. 

\begin{figure}
\begin{subfigure}{.5\textwidth}
\centering
\includegraphics[width=\linewidth]{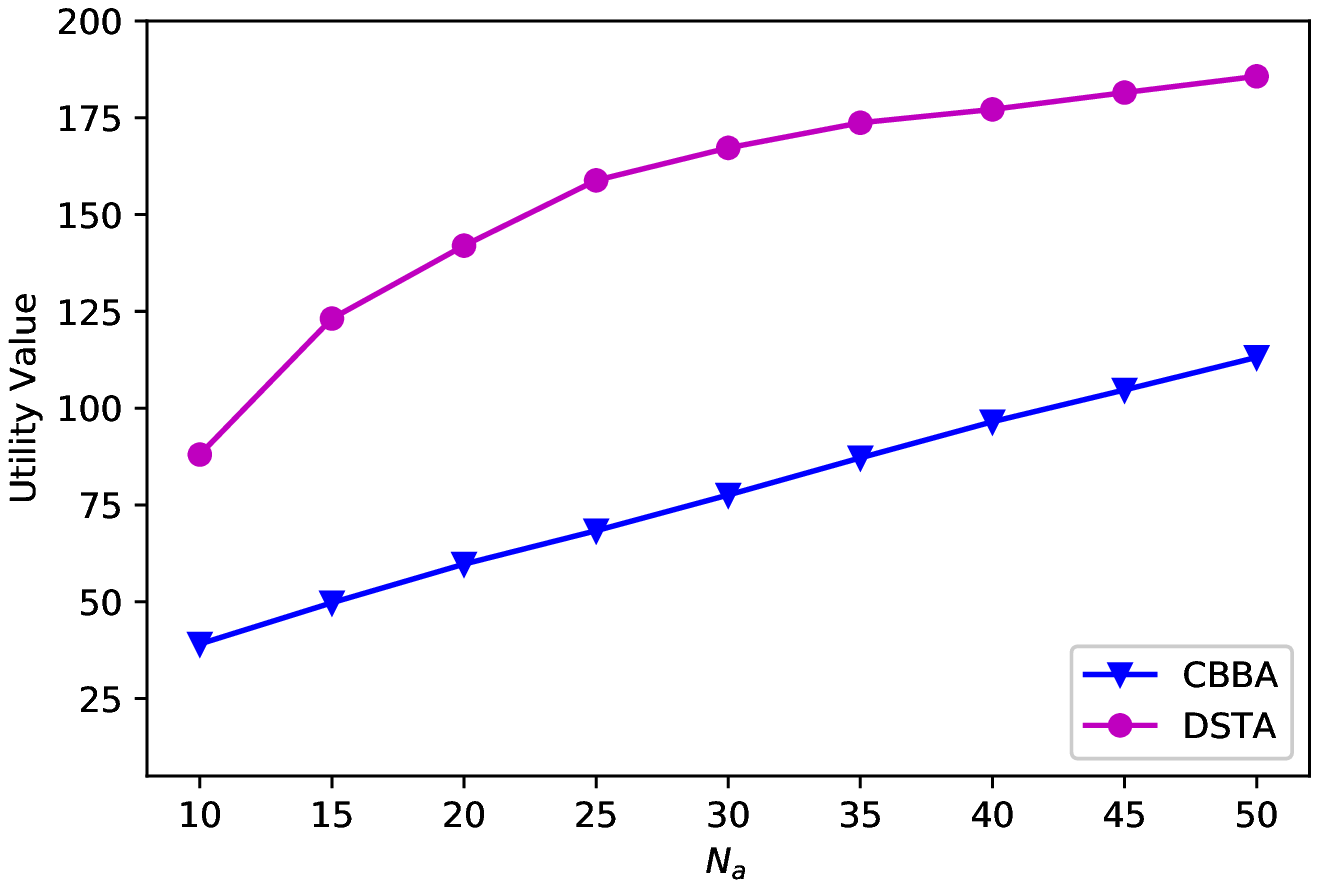}
\caption{Utility function}
\end{subfigure}
\begin{subfigure}{.5\textwidth}
\centering
\includegraphics[width=\linewidth]{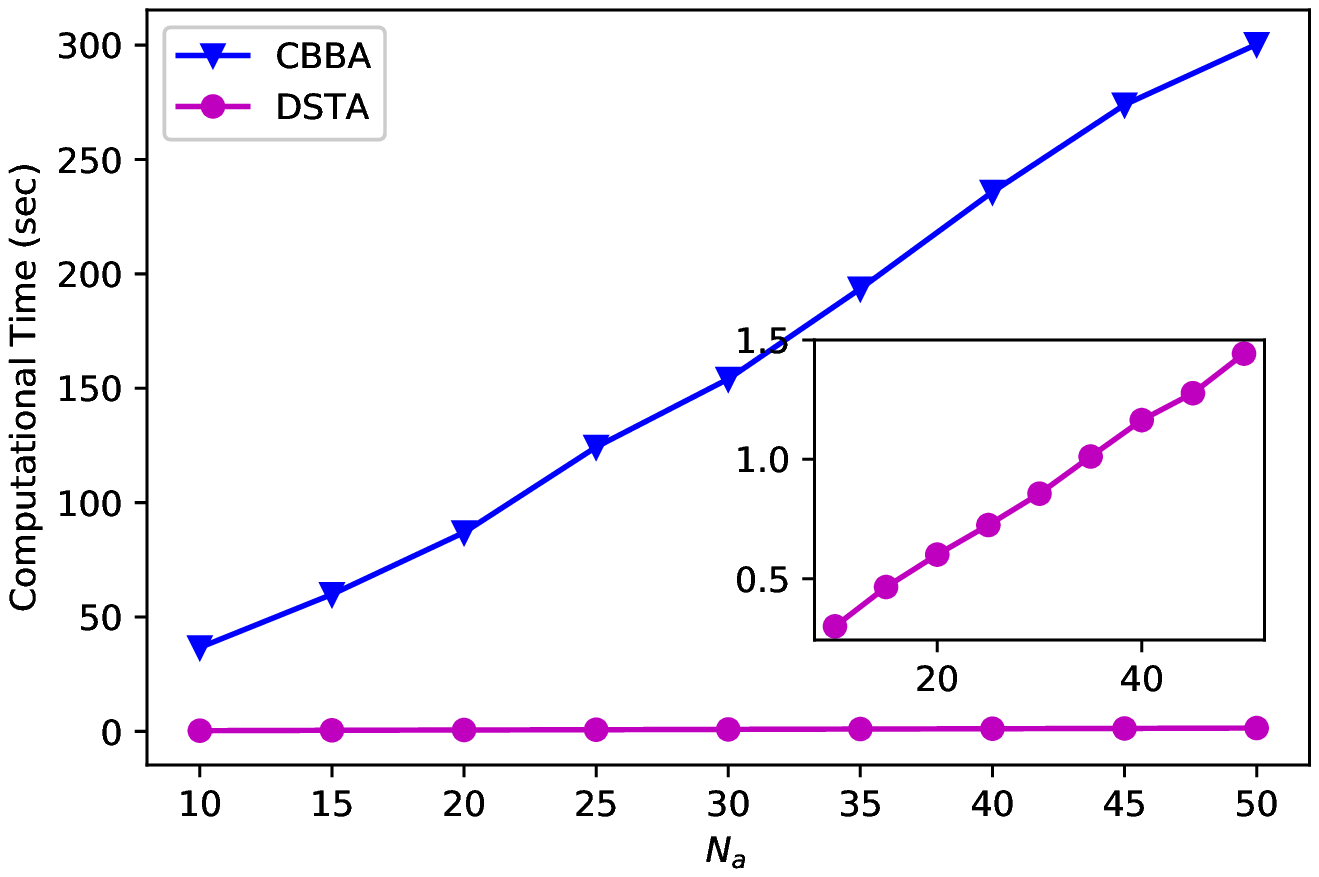}
\caption{Computational time}
\end{subfigure}
\caption{Performance comparison for the non-monotone submodular function ($N_t = 200$)}
\label{fig:non-monotone01}
\end{figure}

\begin{figure}
\begin{subfigure}{.5\textwidth}
\centering
\includegraphics[width=\linewidth]{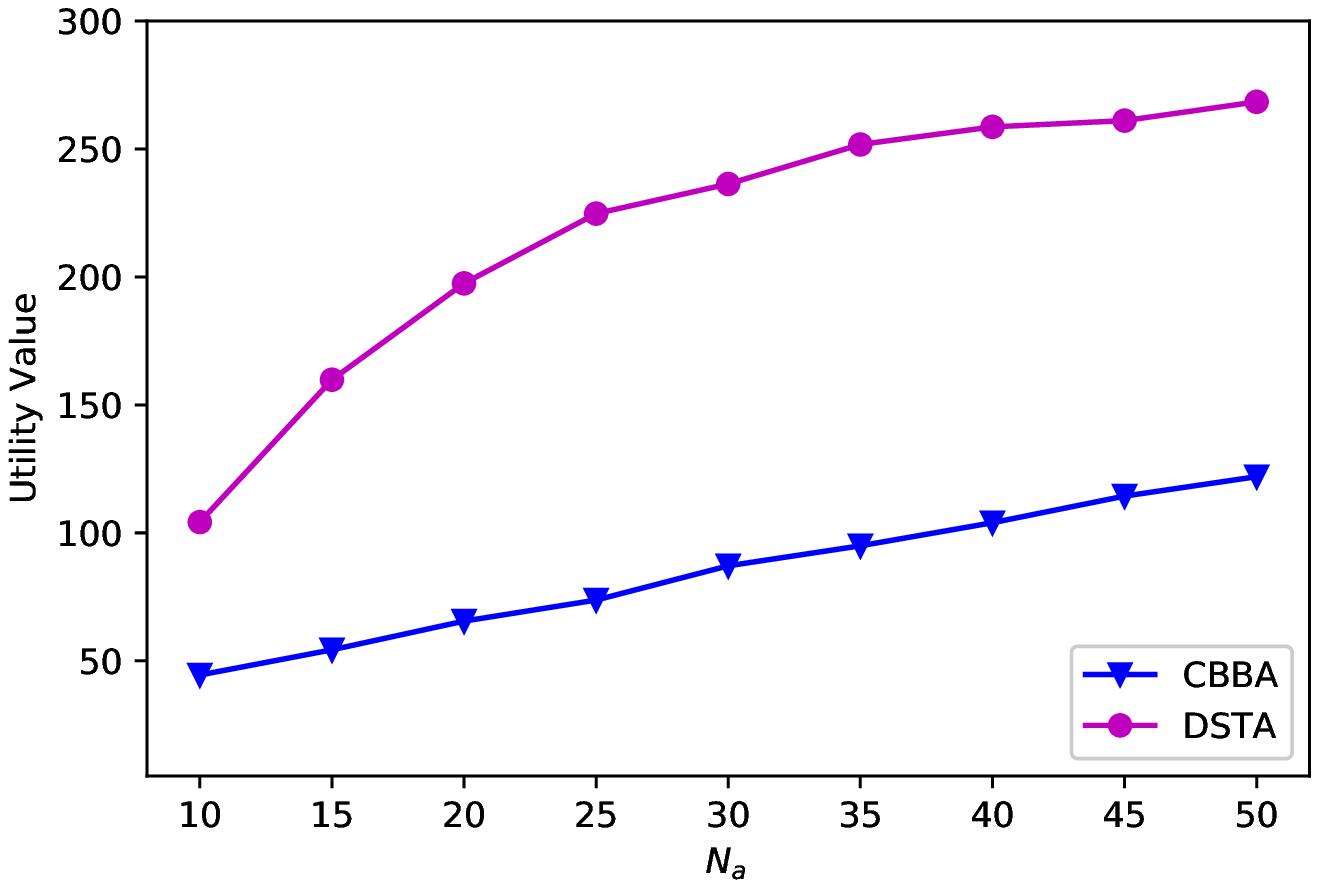}
\caption{Utility function}
\end{subfigure}
\begin{subfigure}{.5\textwidth}
\centering
\includegraphics[width=\linewidth]{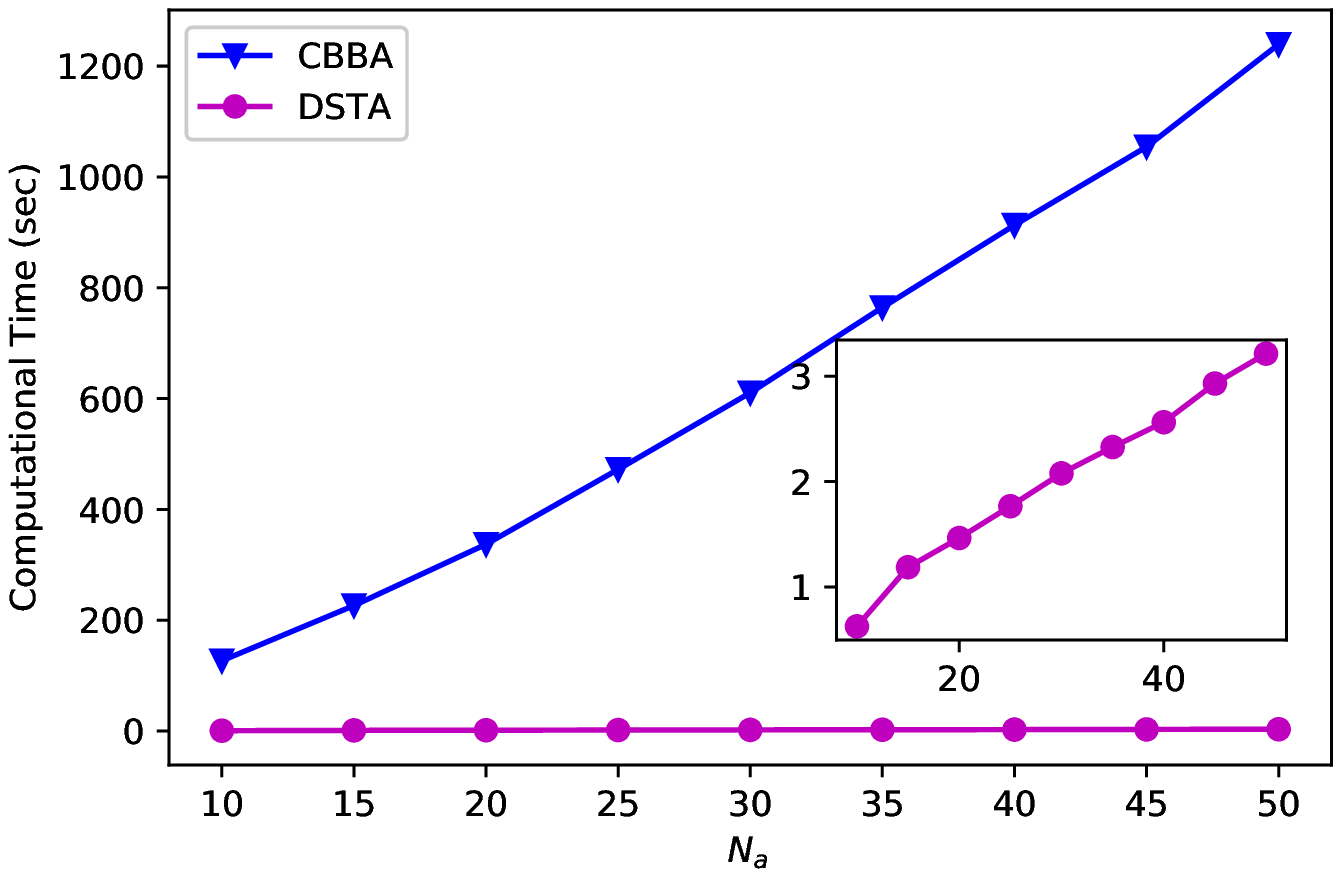}
\caption{Computational time}
\end{subfigure}
\caption{Performance comparison for the non-monotone submodular function ($N_t = 300$)}
\label{fig:non-monotone02}
\end{figure}

The simulation results are depicted in \figs{fig:non-monotone01} and \ref{fig:non-monotone02}. The results confirm that the proposed DSTA achieves much better function values than CBBA: the function values of CBBA achieves around 50\% and 40\% of DSTA for 200 and 300 tasks, respectively. The potential attribute the relatively poor performance of CBBA is that it makes the agents greedily select more important tasks at early iterations. If the agents select more tasks in CBBA, the total function values could start to decrease. Hence, CBBA seems to get trapped in local optima with a few tasks per agent. Note that CBBA only has constant factor guarantees for monotone submodular functions. By contrast, it may exhibit arbitrarily poor performance with non-monotone utility functions, which is confirmed by the simulation results. On the other hand, the sampling procedure in the proposed DSTA might enable abandoning important tasks with a certain probability and hence utilising most of the available tasks to find solutions without getting trapped in local optima.  

Also, the computational time of the proposed is less than that of CBBA. This could be attributed to the fact that CBBA requires reconstruction of task bundles and hence function evaluations whenever there exists a conflict. If the number of conflicts increases, the consensus strategy in CBBA might require more steps for conflict resolution, which results in increased computational complexity.

\begin{figure}
\begin{subfigure}{.5\textwidth}
\centering
\includegraphics[width=\linewidth]{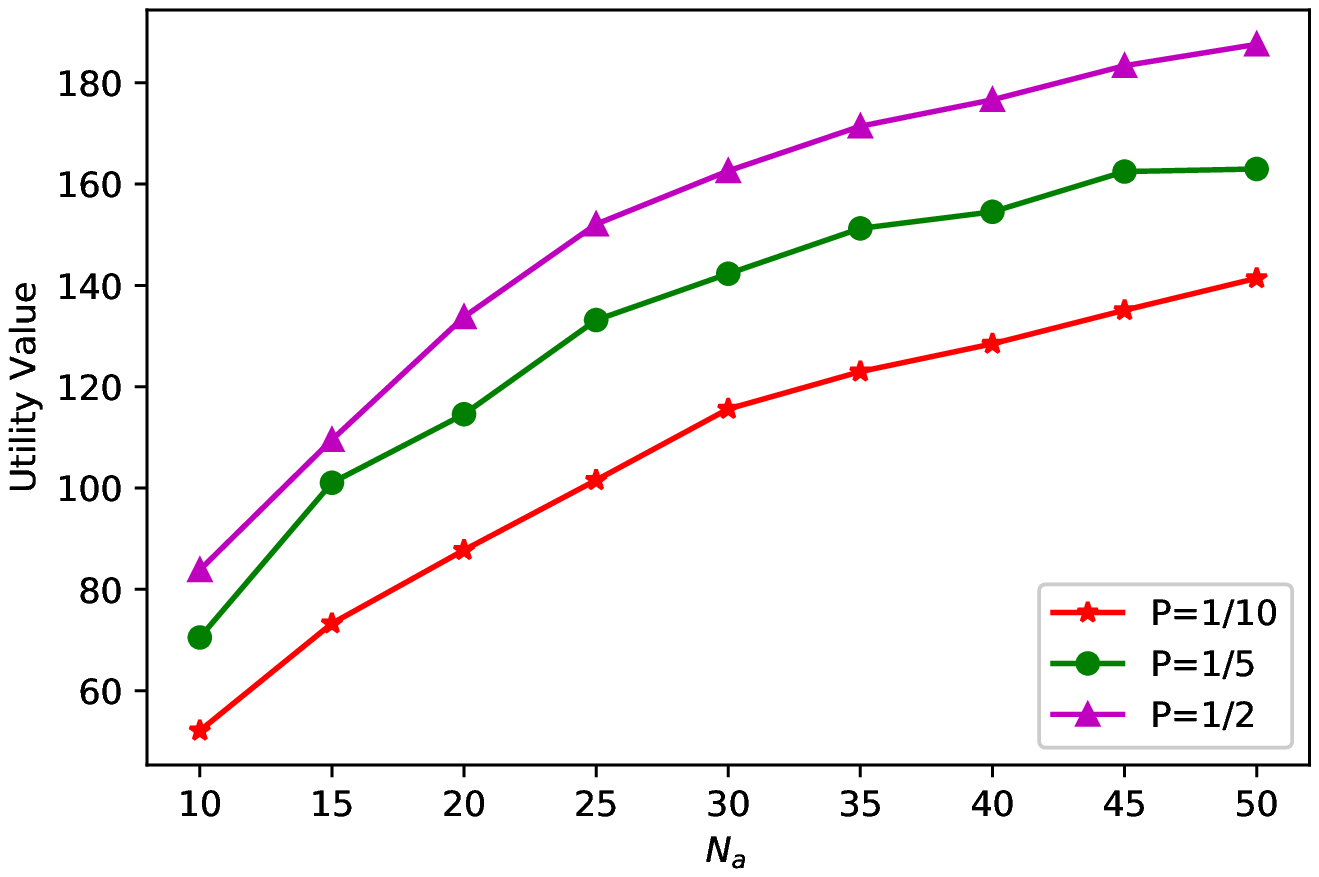}
\caption{Utility function}
\end{subfigure}
\begin{subfigure}{.5\textwidth}
\centering
\includegraphics[width=\linewidth]{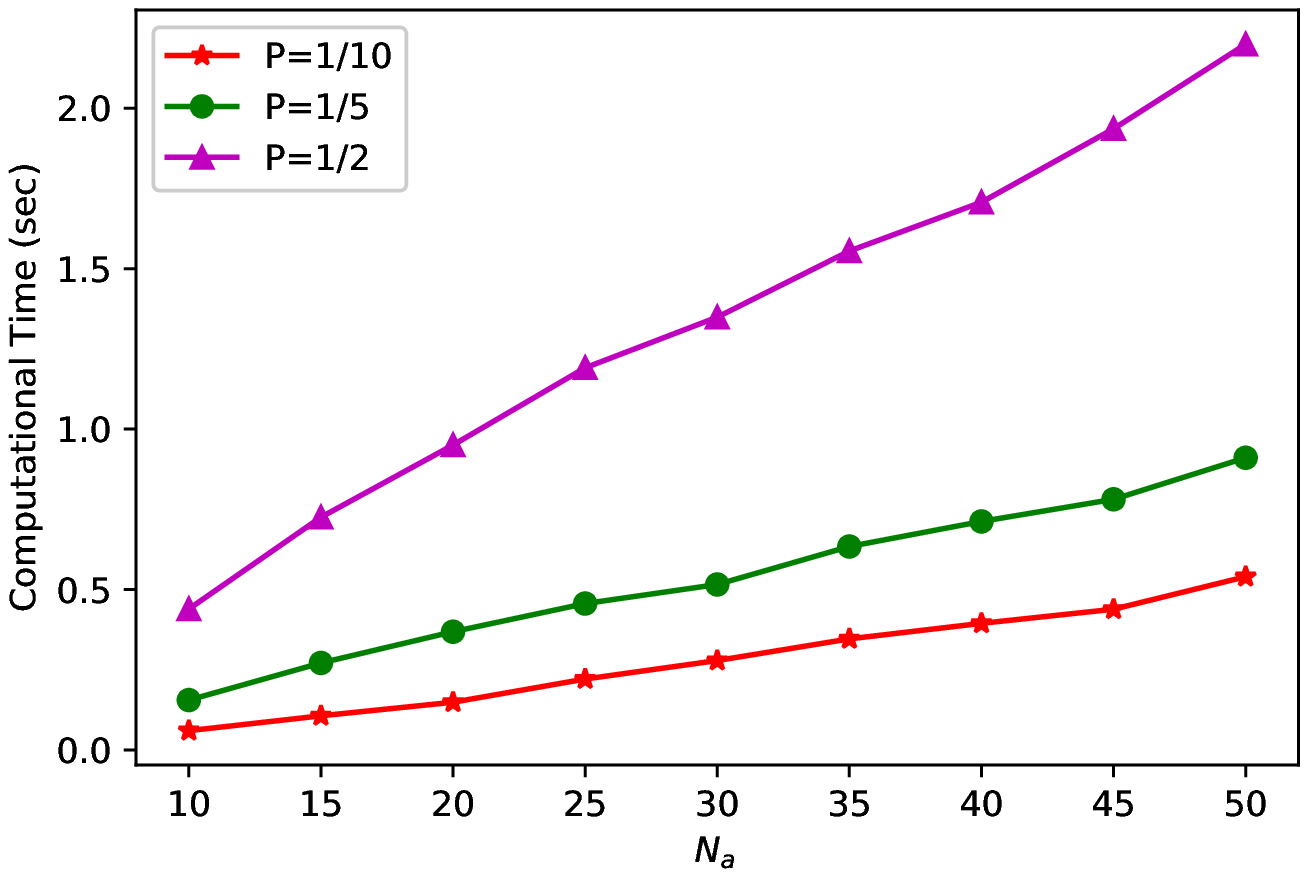}
\caption{Computational time}
\end{subfigure}
\caption{Performance of the DSTA algorithm with respect to different sampling probabilities}
\label{fig:non-monotone_Pcom}
\end{figure}

To validate the analysis results in Section \ref{sec:3}, we also run another set of simulations  with respect to different sampling probabilities for the DSTA algorithm. The sampling probability is set to $1/10$, $1/5$ and $1/2$, respectively, for 200 tasks. \fig{fig:non-monotone_Pcom} shows the simulation results. As demonstrated in \fig{fig:non-monotone_Pcom}, the quality of solutions improves, but the computational time also increases as the sampling probability increases to $1/2$. These results coincide with the analysis results in Section \ref{sec:3} and hence confirm the analysis results. 

%

\section{Conclusions and Future Work}
\label{sec:5}

This paper presents an efficient decentralised task allocation algorithm for \gls{mrs}. Considering the task allocation problem can be viewed as optimisation of a set function subject to a matroid constraint, we leveraged the submodular maximisation concepts for the theoretical tractability. The CBBA algorithm, which is one of the most applied and practical task allocation algorithms, also utilises the submodularity concept and hence provides an approximation guarantee. The issue is that it only provides an approximation guarantee for monotone submodular utility functions. To overcome this issue, this paper utilises a sampling process, i.e. drawing task samples from the set of all tasks. Consequently, the proposed task allocation algorithm achieves an expected approximation guarantee not only for monotone submodular utility functions, but also for general non-monotone submodular utility functions. Moreover, the computational complexity of the proposed algorithm can be further relaxed and adjusted as introduction of the sampling process allows reduction of function evaluations: it requires to evaluate function values for only task samples, not for all the tasks. The performance of the proposed task allocation algorithm is investigated through theoretical analysis. The results of numerical simulations conform the validity of the theoretical analysis results. 

A future research direction would be improving the approximation ratio for both monotone and non-monotone submodular utility functions. There are some gaps between theoretically achievable approximation guarantee and that of our algorithm. The key challenge will be how to improve the approximation guarantee for both monotone and non-monotone cases while maintaining reasonable computational complexity. Another research direction would be further relaxing the computational complexity while guaranteeing the same or similar approximation guarantee. In our opinion, this could be achieved by introducing the lazy greedy concept to the proposed algorithm.

\bibliographystyle{IEEEtran}

\providecommand{\url}[1]{#1} \csname url@samestyle\endcsname%
\providecommand{\newblock}{\relax} \providecommand{\bibinfo}[2]{#2} %
\providecommand{\BIBentrySTDinterwordspacing}{\spaceskip=0pt\relax} %
\providecommand{\BIBentryALTinterwordstretchfactor}{4}
\providecommand{\BIBentryALTinterwordspacing}{\spaceskip=\fontdimen2\font plus
\BIBentryALTinterwordstretchfactor\fontdimen3\font minus
  \fontdimen4\font\relax}
\providecommand{\BIBforeignlanguage}[2]{{\expandafter\ifx\csname l@#1\endcsname\relax
\typeout{** WARNING: IEEEtran.bst: No hyphenation pattern has been}\typeout{** loaded for the language `#1'. Using the pattern for}\typeout{** the default language instead.}\else
\language=\csname l@#1\endcsname
\fi
#2}} \providecommand{\BIBdecl}{\relax} \BIBdecl

\bibliography{submodular_TA}

\end{document}